\documentstyle[11pt]{article}
\begin{document}
\newcommand{\bra}[1]{\langle #1 |}
\newcommand{\ket}[1]{| #1 \rangle}
\unitlength1cm
\jot10pt
\vskip 0.5cm
\begin{flushright}
DTP/95/25\\
quant-ph/9505017
\end{flushright}
\vskip 0.5cm
\begin{center}
{\LARGE Realism and Time Symmetry in Quantum\\
Mechanics\\}
\renewcommand{\thefootnote}{\fnsymbol{footnote}}
\vspace{3.5ex}
{\large Benedikt Bl\"asi\footnote{E-mail: b.m.blasi@durham.ac.uk} and
Lucien Hardy\footnote{E-mail: lucien.hardy@durham.ac.uk} \\[0.8ex]
{\small\it Department of Mathematical Sciences}
\\ {\small\it University of Durham} \\ {\small\it South Road} \\
{\small\it Durham DH1 3LE}}
\renewcommand{\thefootnote}{\arabic{footnote}}
\setcounter{footnote}{0}
\end{center}
\begin{abstract}
We describe a gedanken experiment with an interferometer in the case
of pre- and postselection in two different time symmetric ways:  We
apply the ABL formalism and the de Broglie--Bohm model. Interpreting
these descriptions ontologically, we get two very different concepts
of reality. Finally, we discuss some problems implied by these concepts.
\end{abstract}
\section{Introduction}
\indent

If we are talking about the question of time symmetry/asymmetry in
quantum mechanics, the following statement is quite common: The theory
is time symmetric as long as it can be described by the evolution of a
state vector according to the Schr\"odinger equation. But as soon as
measurement and wave function collapse are involved, the symmetry
breaks down. Time symmetry in quantum mechanics and related topics
are investigated in a number of papers (\cite{abl} -- \cite{deco}).

Aharonov, Bergman and Lebowitz (ABL) \cite{abl} invented a time
symmetric formalism for describing quantum systems between two complete
measurements. This formalism was later generalised by Aharonov and
Vaidman \cite{gen}. The basic idea can be seen in a very simple
experiment: (The set-up is taken from ref. \cite{between}.) Let
$\sigma_x$ and $\sigma_n$ be spin observables corresponding to the
components of a $\mbox{spin-} \frac{1}{2}$ particle along the unit
vectors \boldmath $x$ and $n$. \unboldmath The free Hamiltonian of the
system shall be zero. Let us now consider a particle prepared in the state
$|\sigma_x = \frac{1}{2} \rangle$ at time $t_1$ and found in the state
$\langle \sigma_n = \frac{1}{2} |$ at a later time $t_2$. Obviously, for
an intermediate measurement we get $\mbox{prob}(\sigma_x = \frac{1}{2})
= \mbox{prob}(\sigma_n = \frac{1}{2}) = 1$. Refs. \cite{gen} and
\cite{abl} describe such a pre- and postselected quantum system by two
state vectors (one of them forward and the other backward evolved) and
provide a formula that yields probabilities for outcomes of
intermediate measurements. This formalism is entirely time symmetric.

While in the standard approach the particle in the given example is
completely described by the forward evolved vector $|\sigma_x =
\frac{1}{2} \rangle$
between the two measurements, this time symmetric formalism suggests
that additional information about the intermediate state can be
obtained from the result of the second measurement: In a generalised
state between $t_1$ and $t_2$, the backward evolved vector $\langle
\sigma_n = \frac{1}{2} |$ is also taken into account. Therefore, one
could wonder whether the described particle in the intermediate state
somehow has fixed values for the spin in two different directions.
More generally, this raises the question, of whether the time symmetric
description allows any conclusions about the ontology of a pre- and
postselected quantum system. In terms of causality, such an ontology
could be counterintuitive (cf. \cite{arrow}). While in a deterministic
world (as in classical mechanics or the de Broglie--Bohm model) time
symmetry and causality are in a `peaceful
coexistence'\footnote{A. Shimony has used the same phrase with respect to
nonlocality in quantum mechanics and the impossibility of superluminal
signaling. (cf. ref. \cite{peace})}, in the
probabilistic interpretation of quantum mechanics there can appear a
problem with time symmetry: Timelike correlations between measurements
could imply something like `precognitive elements' in the quantum system.

In this paper, we shall describe an interferometer experiment with
pre- and postselection (analogous to the above mentioned set-up) and
apply the time symmetric formalism to it. Furthermore, we will
investigate possible consequences of this formalism for the ontology
or elements of reality in the described system. Finally, we shall
compare these results with the de Broglie--Bohm model as a realistic
interpretation of quantum mechanics.
\\
\section{Description of the experimental set-up \label{ex}}
\indent

We will consider an interferometer as shown in Fig.~1 with a single
particle source. It shall be arranged so that different paths
between two beamsplitters (i.e. $c$ and $d$, $e$ and $f$) have exactly
the same length.
%
\begin{figure}[bt]
\begin{center}
\begin{picture}(9,10)
\put(0,1){
\begin{picture}(9,9)
\put(0,1){\vector(1,0){0.5}}
\put(0.5,1){\vector(1,0){2}}
\put(2.5,1){\line(1,0){1.5}}
\put(1,4){\vector(1,0){1.5}}
\put(2.5,4){\vector(1,0){3}}
\put(5.5,4){\line(1,0){1.5}}
\put(4,7){\vector(1,0){1.5}}
\put(5.5,7){\vector(1,0){2}}
\put(7.5,7){\line(1,0){0.5}}
\put(1,0){\vector(0,1){0.5}}
\put(1,0.5){\vector(0,1){2}}
\put(1,2.5){\line(0,1){1.5}}
\put(4,1){\vector(0,1){1.5}}
\put(4,2.5){\vector(0,1){3}}
\put(4,5.5){\line(0,1){1.5}}
\put(7,4){\vector(0,1){1.5}}
\put(7,5.5){\vector(0,1){2}}
\put(7,7.5){\line(0,1){0.5}}
\multiput(0.5,0.5)(3,3){3}{\line(1,1){1}}
\put(1.7,1.7){\makebox(0,0){\large BS 1}}
\put(4.7,4.7){\makebox(0,0){\large BS 2}}
\put(6.3,6.3){\makebox(0,0){\large BS 3}}
\multiput(3.5,0.5)(3,3){2}{\line(1,1){1}}
\multiput(3.5,0,5)(0.2,0.2){6}{\line(0,-1){0.15}}
\multiput(6.5,3,5)(0.2,0.2){6}{\line(0,-1){0.15}}
\multiput(0.5,3.5)(3,3){2}{\line(1,1){1}}
\multiput(0.5,3,5)(0.2,0.2){6}{\line(0,1){0.15}}
\multiput(3.5,6,5)(0.2,0.2){6}{\line(0,1){0.15}}
\put(0.5,1.3){\makebox(0,0){\large a}}
\put(2.5,4.3){\makebox(0,0){\large d}}
\put(5.5,7.3){\makebox(0,0){\large f}}
\put(1.3,0.5){\makebox(0,0){\large b}}
\put(2.5,0.7){\makebox(0,0){\large c}}
\put(5.5,3.7){\makebox(0,0){\large e}}
\put(7.5,6.7){\makebox(0,0){\large g}}
\put(6.7,7.5){\makebox(0,0){\large h}}
\put(8,6.6){\framebox(0.8,0.8){\Large G}}
\put(6.6,8){\framebox(0.8,0.8){\Large H}}
\put(0,-0.5){\parbox[t]{9cm}{\footnotesize Fig.~1}}
\end{picture}
}
\end{picture}
\end{center}
\end{figure}
As the following calculations show, the set-up is composed of two
balanced Mach--Zehnder type interferometers. This means that a single
wave, incoming from path $a$ ($b$) leads to a single wave in path $e$
($f$). And if at $BS\ 2$ there approaches a wave only from path $c$
($d$), then detector $H$ ($G$) will `fire' with probability 1.

The operation of every beamsplitter $BS$ (see Fig.~2) on the state
vectors $|u \rangle$ and $|v \rangle$ is given by
%
\begin{eqnarray}
|u\rangle & \stackrel{BS}{\longrightarrow} &
  \frac{1}{\sqrt{2}} (|x\rangle + i|y\rangle) \label{u} \\
|v\rangle & \stackrel{BS}{\longrightarrow} &
  \frac{1}{\sqrt{2}} (i|x\rangle + |y\rangle) . \label{v}
\end{eqnarray}
%
\begin{figure}[ht]
\begin{center}
\begin{picture}(2,3.8)
\put(0,1){
\begin{picture}(2,2.8)
\put(0,1){\vector(1,0){0.5}}
\put(0.5,1){\vector(1,0){1}}
\put(1.5,1){\line(1,0){0.5}}
\put(1,0){\vector(0,1){0.5}}
\put(1,0.5){\vector(0,1){1}}
\put(1,1.5){\line(0,1){0.5}}
\put(0.5,0.5){\line(1,1){1}}
\put(-0.3,1){\makebox(0,0){\large u}}
\put(1,2.3){\makebox(0,0){\large y}}
\put(1,-0.3){\makebox(0,0){\large v}}
\put(2.3,1){\makebox(0,0){\large x}}
\put(1.7,1.7){\makebox(0,0){\large BS}}
\put(-0.4,-0.8){\parbox[t]{2cm}{\footnotesize Fig. 2}}
\end{picture}
}
\end{picture}
\end{center}
\end{figure}

\noindent
Using these conditions, the evolution of an initial state $|a \rangle$
is given by
%
\begin{eqnarray}
|a\rangle & \stackrel{BS\:1}{\longrightarrow} &
  \frac{1}{\sqrt{2}} (|c\rangle + i|d\rangle)
  \qquad (\equiv |\psi_1 \rangle) \label{for1}\\
& \stackrel{BS\:2}{\longrightarrow} &
  \frac{1}{\sqrt{2}} (\frac{1}{\sqrt{2}} (i|e\rangle + |f\rangle)
  + \frac{i}{\sqrt{2}} (|e\rangle + i|f\rangle)) \nonumber \\
& & = i|e\rangle \label{for2}\\
& \stackrel{BS\:3}{\longrightarrow} &
  \frac{i}{\sqrt{2}} (i|g\rangle + |h\rangle) \nonumber \\
& & = \frac{1}{\sqrt{2}} (-|g\rangle + i|h\rangle) . \label{for3}
\end{eqnarray}
\\
\section{Element of reality in the case of pre- and postselection \label{sel}}
\indent

We will now recall the time symmetric description of pre- and
postselected quantum systems that was first invented by Aharonov,
Bergman and Lebowitz (ABL) in 1964 \cite{abl}. Here, we shall use the
generalised formalism as introduced by Aharonov and Vaidman
\cite{gen}.

Let us consider a quantum system  prepared in a state $|\psi_1 (t_1)
\rangle$ at time $t_1$ and postselected in a state $\langle \psi_2
(t_2)|$ at a later time $t_2$. By using the time evolution operator
$U$, we get a forward evolved state vector
\[ |\psi_1 (t) \rangle = U(t_1,t) |\psi_1 (t_1) \rangle \]
as well as a backward evolved state (denoted as a `bra'-vector)
\[ \langle \psi_2 (t)| = \langle \psi_2 (t_2)| U(t,t_2) \]
for any time $t$ with $t_1 < t < t_2$. With this, the generalised
state at that time is defined as a vector that includes both the
backward evolved state (`bra') and the forward evolved state
(`ket')\footnote{To prevent confusion, it should be
emphasized that this term is {\it not} a scalar product!}:
\begin{equation}
{\bf \Psi} (t) \equiv \langle \psi_2 (t)||\psi_1 (t) \rangle
\label{gen}
\end{equation}
And the probability that an intermediate measurement of an operator
$C$ at time $t$ yields the eigenvalue $c_n$ is given by the ABL
formula
\begin{equation}
\mbox{prob}(C=c_n)= \frac{|\langle \psi_2 (t)| P_{C=c_n} |\psi_1 (t)
\rangle|^2}{\sum_i |\langle \psi_2 (t)| P_{C=c_i} |\psi_1 (t)
\rangle|^2} \; ,
\label{abl}
\end{equation}
where $P_{C=c_i}$ is the projection operator on the space of
eigenstates with eigenvalue $c_i$.

Let us now consider the set-up of section \ref{ex} only in situations
where the initial state is $|a \rangle$ and where at the end the
particle is detected at $G$. Consequently, between preparation and
detection we have a pre- and postselected quantum system and we can
employ the above formalism.

The backward evolution at a beamsplitter (see Fig.~2) can be obtained
{}from (\ref{u}) and (\ref{v}) in a straightforward calculation:
%
\begin{eqnarray}
\langle x| & \stackrel{BS}{\longrightarrow} &
  \frac{1}{\sqrt{2}} (\langle u| - i\langle v|) \\
\langle y| & \stackrel{BS}{\longrightarrow} &
  \frac{1}{\sqrt{2}} (-i\langle u| + \langle v|)
\end{eqnarray}
And so, the postselected state $\langle g|$ evolves backwards as
follows:
%
\begin{eqnarray}
\langle g| & \stackrel{BS\:3}{\longrightarrow} &
  \frac{1}{\sqrt{2}} (\langle f| - i\langle e|) \label{back1} \\
& \stackrel{BS\:2}{\longrightarrow} & -i\langle d|
  \qquad (\equiv \langle \psi_2|) \label{back2} \\
& \stackrel{BS\:1}{\longrightarrow} &
  -\frac{1}{\sqrt{2}} (\langle a| + i\langle b|) \label{back3}
\end{eqnarray}
Therefore, as the generalised state (cf. (\ref{gen})) between $BS\;1$
and $BS\;2$ we obtain
%
\begin{equation}
\langle \psi_2||\psi_1 \rangle  =
  \frac{1}{\sqrt{2}} \langle d| \; (-i|c \rangle + |d \rangle).
\end{equation}

We shall now imagine the detection of the particle between $BS\;1$ and
$BS\;2$ as an intermediate measurement with observable $D$. Possible
results are $D=0$ (detection in path $c$) and $D=1$ (detection in path
$d$). So the ABL formula (cf. (\ref{abl})) yields the probability of
detecting the particle in path $d$:
%
\begin{equation}
\mbox{prob}(D=1)=\frac{|\langle \psi_2|d \rangle \langle d|\psi_1 \rangle|^2}
  {|\langle \psi_2|c \rangle \langle c|\psi_1 \rangle|^2
   + |\langle \psi_2|d \rangle \langle d|\psi_1 \rangle|^2} = 1 .
\label{prob}
\end{equation}
An analogous calculation yields that in a similar measurement between
$BS\;2$ and $BS\;3$ the particle would be detected in path $e$ with
probability 1.
%
\begin{figure}[bt]
\begin{center}
\begin{picture}(9,10.5)
\put(0,1.5){
\begin{picture}(9,9)
\put(1,0){\line(0,1){1}}
\put(1,1){\line(1,0){3}}
\put(4,1){\line(0,1){6}}
\put(4,7){\line(1,0){3}}
\put(7,7){\line(0,1){1}}
\linethickness{0.6mm}
\put(0,1){\line(1,0){1}}
\put(1,1){\line(0,1){3}}
\put(1,4){\line(1,0){6}}
\put(7,7){\line(1,0){1}}
\put(7,4){\line(0,1){3}}
\thinlines
\multiput(0.5,0.5)(3,3){3}{\line(1,1){1}}
\put(1.7,1.7){\makebox(0,0){\large BS 1}}
\put(4.7,4.7){\makebox(0,0){\large BS 2}}
\put(6.3,6.3){\makebox(0,0){\large BS 3}}
\multiput(3.5,0.5)(3,3){2}{\line(1,1){1}}
\multiput(3.5,0,5)(0.2,0.2){6}{\line(0,-1){0.15}}
\multiput(6.5,3,5)(0.2,0.2){6}{\line(0,-1){0.15}}
\multiput(0.5,3.5)(3,3){2}{\line(1,1){1}}
\multiput(0.5,3,5)(0.2,0.2){6}{\line(0,1){0.15}}
\multiput(3.5,6,5)(0.2,0.2){6}{\line(0,1){0.15}}
\put(0.5,1.3){\makebox(0,0){\large a}}
\put(2.5,4.3){\makebox(0,0){\large d}}
\put(5.5,7.3){\makebox(0,0){\large f}}
\put(1.3,0.5){\makebox(0,0){\large b}}
\put(2.5,0.7){\makebox(0,0){\large c}}
\put(5.5,3.7){\makebox(0,0){\large e}}
\put(7.5,6.7){\makebox(0,0){\large g}}
\put(6.7,7.5){\makebox(0,0){\large h}}
\put(8,6.6){\framebox(0.8,0.8){\Large G}}
\put(6.6,8){\framebox(0.8,0.8){\Large H}}
\put(0,-0.5){\parbox[t]{9cm}{\footnotesize Fig.~3: Paths of the
interferometer where for a particle
preselected in state $|a \rangle$ and postselected in state $\langle
g|$ the time symmetric description suggests elements of reality}}
\end{picture}
}
\end{picture}
\end{center}
\end{figure}

This result of a so far purely formal description now could be
interpreted ontologically.
Let us therefore  recall Redhead's \cite{real} {\it ``Sufficient Condition
for Element of Reality''}. (Originally, this condition was used in the
EPR argument.) He states: {\it ``If we can predict with
certainty, or at any rate with probability one the result of measuring
a physical quantity at time t, then at the time t, there exists an
element  of reality corresponding to this physical quantity and having
a value equal to the predicted measurement result.''} As a
modification of this, Vaidman \cite{mreal} suggests to replace {\it
predict} by {\it infer}. So, unlike Redhead's condition, the
statement would no longer be time biased.

Encouraged by these definitions, with our above result (\ref{prob})
one could conclude that it is an element of physical reality that in
our gedanken experiment the particle goes through path $d$ and $e$
(see Fig. 3).

It should be mentioned that, apart from this ontological
interpretation of the ABL formula, the time symmetric formalism can
lead to other statements about an underlying reality, too. In
particular, it should be worth thinking about a possible ontological
meaning of the generalised state (\ref{gen}).

But all the concepts of reality based on the above time
symmetric description obviously have one complication in common:
Because these ontologies depend on the
performance of a particular final measurement and on its outcome,
there appear to be contradictions with causality.
\\
\section{Time symmetric description of a measurement\label{meas}}
\indent

A subtle point of the above mentioned time symmetric description is
the measurement process. In particular, one could ask how time biased
concepts such as detection or outcome can be compatible with time
symmetry.

To see this, let us consider a measurement of the von Neumann
\cite{neu} type: If we want to measure an observable $A$ of a quantum
system $S$, then the interaction Hamiltonian between $S$ and a
measurement apparatus $M$ is given by
\begin{equation} H_{int} = g(t)pA . \end{equation}
The normalised coupling function $g(t)$ shall be nonzero for a short
time interval. The momentum $p$ is the canonical conjugate to a
pointer position $q$. For simplicity, we assume the free Hamiltonian
to equal zero.

Let the initial states of $S$ and $M$ be $\ket{\psi_i}_s$ and
$\ket{\phi_i}_m$, respectively. If we denote the eigenstates of $A$
by $\ket{a_k}_s$ (with $A \ket{a_k}_s = a_k \ket{a_k}_s$), then
$\ket{\psi_i}_s$ can be expanded as follows: $\ket{\psi_i}_s = \sum_k
\alpha_k \ket{a_k}_s$. Position eigenstates of $M$ are denoted by
$\ket{q}_m$.

The forward evolution of $\ket{\psi_i}_s \ket{\phi_i}_m$ during the
measurement can be described in three steps:\\
\begin{tabular}{rl}
(i) & preparation of $M$ (reading of the pointer position $q_1$, collapse
1)\\
(ii) & measurement interaction\\
(iii) & reading the result (the pointer position $q_2$, collapse 2)
\end{tabular}
\begin{eqnarray}
\ket{\psi_i}_s \ket{\phi_i}_m & \stackrel{coll. 1}{\longrightarrow} &
  \ket{\psi_i}_s \ket{q_1}_m \\
& \stackrel{H_{int}}{\longrightarrow} &
  \sum_k \alpha_k \ket{a_k}_s \ket{q_1+a_k}_m \\
& \stackrel{coll. 2}{\longrightarrow} & \ket{a_l}_s \ket{q_1+a_l}_m
\end{eqnarray}
The final reading yields
\begin{equation} q_2 = q_1 + a_l , \label{fir} \end{equation}
and hence, we can deduce the result, $a_l$, of the measurement from a
knowledge of $q_1$ and $q_2$.

Let us now consider this process in the reversed time direction. In this
case, the evolution `starts' in a final state $\bra{\psi_f}_s
\bra{\phi_f}_m$ with $\bra{\psi_f}_s = \sum_k \beta_k \bra{a_k}_s$.
Here, the observer prepares $M$ by reading the position $q_2$
(collapse 2) and reads the result $q_1$ `afterwards' (collapse 1):
\begin{eqnarray}
\bra{\psi_f}_s \bra{\phi_f}_m & \stackrel{coll. 2}{\longrightarrow} &
  \bra{\psi_f}_s \bra{q_2}_m\\
& \stackrel{H_{int}}{\longrightarrow} &
  \sum_k \beta_k \bra{a_k}_s \bra{q_2 - a_k}_m\\
& \stackrel{coll. 1}{\longrightarrow} & \bra{a_n}_s \bra{q_2 - a_n}_m
\end{eqnarray}
In this case, the result is
\begin{equation} q_1 = q_2 - a_n , \label{sec} \end{equation}
and again, we can get the measurement result, $a_n$, if we know $q_1$
and $q_2$.

It follows from equns. (\ref{fir}) and (\ref{sec}) that $a_l = a_n$,
and hence, the result of the measurement as deduced by the forward
time observer is the same as that deduced by the backward time
observer. The reason for this is that it is not just the `final'
pointer reading that determines the measurement result, but rather the
{\it difference} between the initial and final pointer reading.
The measurement process is
therefore described in an entirely time symmetric way.
\\
\section{The same experiment, described in the de Broglie--Bohm model
\label{bohm}}
\indent

Considering elements of reality, it shall be interesting to look at a
realistic interpretation of quantum mechanics. The de Broglie--Bohm
model \cite{bohm} provides such an interpretation. It describes all
processes in a time symmetric and deterministic way. (In ref.
\cite{dgz}, the derivation of Bohmian mechanics is even based on time
symmetry.) Knowing the whole
configuration of a system at one arbitrary instant, one can therefore
exactly describe the system at any other instant by backward or forward
evolution. And because in this interpretation all particles have
definite positions at every time, it provides a clear ontology.
As long as $\rho = |\psi|^2 $ initially, this remains true at
all later times. Consequently, the de Broglie--Bohm model has exactly
the same predictions as standard quantum mechanics. This means in
particular that the uncertainty relations hold (cf. \cite{dgz}).
Nevertheless, in the following we will show that
in our experiment of section \ref{ex} the outcome of the detection  at
$G$ or $H$ enables us to state which path the particle went according
to the de Broglie--Bohm model.

So let us again look at the interferometer set-up with the initial
state $|a \rangle$. Dewdney \cite{tra} has shown how in the de
Broglie--Bohm interpretation we can describe trajectories of particles
(guided by wave packets) at beamsplitters
and mirrors: If a wave arrives at a beamsplitter only from one
path, a particle with position in the trailing (leading) half of the
wave packet will be (will not be) reflected. If a wave packet is reflected,
then the order of the particles will be reversed afterwards. (This can
be derived using the
fact that Bohm trajectories are unique and cannot intersect each other
in spacetime.)  According to that, at $BS\:1$ (cf. (\ref{for1})) the
particle goes along path $c$ ($d$), if it adopted a position within
the the front (rear) half of the initial wave packet. Analogous, at
$BS\:2$ (cf. (\ref{for1}) -- (\ref{for2})) particles coming from path
$c$ ($d$) end up in the leading (trailing) half of the wave packet in path
$e$. (Same length of paths $c$ and $d$ is needed here.) As an effect
of the mirror in this path, the order of the particle positions within
the wave packet gets reversed. Finally, $BS\:3$ (cf. (\ref{for2}) --
(\ref{for3})) operates so that a particle detected at $G$ ($H$)
adopted a position within the rear (front) half of the arriving wave
packet. Computing all these steps, we get the result that, according
to the de Broglie--Bohm interpretation, a particle preselected in
state $|a \rangle$ and postselected in state $\langle g|$ between
$BS\:1$ and $BS\:2$
always\footnote{{\it Always} means in the idealised case of our
gedanken experiment {\it with probability 1}; we have not considered
particles with positions exactly in the middle of a wave packet (a set
of measure 0 in position space).} goes through path $c$. (see Fig. 4)
%

\begin{figure}[ht]
\begin{center}
\begin{picture}(9,10)
\put(0,1){
\begin{picture}(9,9)
\put(1,0){\line(0,1){4}}
\put(4,4){\line(0,1){3}}
\put(7,7){\line(0,1){1}}
\put(1,4){\line(1,0){3}}
\put(4,7){\line(1,0){3}}
\linethickness{0.6mm}
\put(0,1){\line(1,0){4}}
\put(4,4){\line(1,0){3}}
\put(7,7){\line(1,0){1}}
\put(4,1){\line(0,1){3}}
\put(7,4){\line(0,1){3}}
\thinlines
\multiput(0.5,0.5)(3,3){3}{\line(1,1){1}}
\put(1.7,1.7){\makebox(0,0){\large BS 1}}
\put(4.7,4.7){\makebox(0,0){\large BS 2}}
\put(6.3,6.3){\makebox(0,0){\large BS 3}}
\multiput(3.5,0.5)(3,3){2}{\line(1,1){1}}
\multiput(3.5,0,5)(0.2,0.2){6}{\line(0,-1){0.15}}
\multiput(6.5,3,5)(0.2,0.2){6}{\line(0,-1){0.15}}
\multiput(0.5,3.5)(3,3){2}{\line(1,1){1}}
\multiput(0.5,3,5)(0.2,0.2){6}{\line(0,1){0.15}}
\multiput(3.5,6,5)(0.2,0.2){6}{\line(0,1){0.15}}
\put(0.5,1.3){\makebox(0,0){\large a}}
\put(2.5,4.3){\makebox(0,0){\large d}}
\put(5.5,7.3){\makebox(0,0){\large f}}
\put(1.3,0.5){\makebox(0,0){\large b}}
\put(2.5,0.7){\makebox(0,0){\large c}}
\put(5.5,3.7){\makebox(0,0){\large e}}
\put(7.5,6.7){\makebox(0,0){\large g}}
\put(6.7,7.5){\makebox(0,0){\large h}}
\put(8,6.6){\framebox(0.8,0.8){\Large G}}
\put(6.6,8){\framebox(0.8,0.8){\Large H}}
\put(0,-0.5){\parbox[t]{9cm}{\footnotesize Fig.~4: Path of a particle
preselected in state $|a \rangle$ and postselected in state $\langle
g|$, according to the de Broglie--Bohm interpretation}}
\end{picture}
}
\end{picture}
\end{center}
\end{figure}
In order to check the above claimed time symmetry of this model, we
will now look at the same experiment in a time reversed sense. That
means, we have a particle `incoming' in path $g$ and `detected' in
$a$. Assuming therefore, that the `initial' state is $\langle g|$ and
the `final' state is $|a \rangle$, with the backward evolution
(\ref{back1}) -- (\ref{back3}) and considerations analogous to them in
the previous paragraph, we obtain a particle trajectory going through
the paths $f$ and $d$ instead of $e$ and $c$. Consequently, to get the
correct `initial state' in the backward picture, it is not sufficient
just to know about the `click' at detector $G$ in the actual
experiment. Additionally, one has to take in account that at the same
time an `empty wave' with a certain phase difference arrives at $H$.
In order to describe the time reversed experiment properly, one
therefore has to evolve the state $\frac{1}{\sqrt{2}} (\langle g| -
i\langle h|)$. In this way, the forward and backward picture yield the
same trajectory.

This makes clear once more that the de Broglie--Bohm model is
dualistic: Both particles and waves are regarded as existing in the
physical world. And a description cannot be complete, if it does not
entirely include both of these entities. Therefore not even `empty
waves' or their phases can be neglected\footnote{The significance of `empty
waves' was already shown in \cite{empty} in an example where
an `empty wave' interacts with a particle. It is quite amusing to see
that one can even get an interaction between two `empty waves', if one
extends the set-up of section \ref{ex} by placing the detection box A
described in \cite{empty} in path $d$. If now the initial state is $|a
\rangle$ and in the end the particle is detected at $G$, then (with
the above considerations about Bohm trajectories) one knows that the
particle went through path $c$ and that path $d$ was not blocked. From
here on, the further conclusions are completely analogous to the
reasoning in \cite{empty}.}. So, in the context of our considerations
about time
symmetry, we can wonder, why we accept an outgoing `empty wave' as
natural, while we usually do not  think of incoming `empty waves'. ---
Could it actually be true that in quantum mechanics there are incoming
`empty waves' as well as outgoing ones? Or could there perhaps be a
freedom of adding `empty waves' to an initial state? --- This
undoubtedly is impossible, because initial states with additional
`empty waves' would lead to (actually not observed) different
outcomes. (For a related discussion see E. J. Squires \cite{hist}.)

Could it therefore be that after all this time asymmetry connected
with the occurrence of `empty waves' only before measurements
indicates a direction of
time inherent in the de Broglie--Bohm model? Because the
dynamical laws in this interpretation are completely time symmetric,
an arrow of time (if there is one) probably cannot be based on a
fundamental level. Another look at Bohm's papers \cite{bohm} shows
that after measurements there {\it are} actually outgoing `empty
waves'. But these waves are entangled with the respective states of
the measurement apparatus. Because as a macroscopical system this
apparatus has a large number of internal degrees of freedom, an
overlap of different waves after the measurement is very unlikely.
Therefore, the probability is neglegibly small that an outgoing `empty
wave' has an influence on the actual position of the system, and for
all practical purposes, we can replace the complete wave function by a
new renormalised one (the prepared state).With this,
the time asymmetry inherent in the measurement process (as
described by Bohm) is `reduced' to the thermodynamical arrow of time.
\\
\section{Conclusions}
\indent

In this paper, we presented two different time symmetric descriptions
of quantum mechanics: the ABL formalism and the de Broglie--Bohm
model. As the interferometer experiment shows, these
interpretations suggest entirely different ontologies. However, we
cannot go as far as to conclude that these models of an underlying
physical reality actually contradict each other. In fact, we are
dealing with two different ontological {\it concepts} that need to be
clarified much more in order to get any statements about mutual
consistency or contradiction.

In section \ref{bohm}, we discuss particle trajectories.
Our result is that in the described set-up we know (according to
Bohm) the path of the particle, if we are informed about initial
preparation and final detection. This path ($c$) is actually different from
the path ($d$) where the ABL formula suggests an element of reality.
However, in section \ref{bohm}, we also emphasized the significance of
`empty waves' in the de Broglie--Bohm model. With such an `empty
wave', this model describes something physically real going
through path $d$, too.
Since particle positions are the only thing we can actually
observe, one often tends to give the particles in the de Broglie--Bohm
interpretation `more reality' than the waves. However, one has to be very
careful with this idea.

For the discussion of the quite formally defined elements of reality
that were suggested in the context of the ABL formalism
in section \ref{sel}, it shall be useful to recall the idea of
the definition: It is based on the probability for a certain outcome
of an intermediate measurement. (Since we are dealing with a pre-
and postselected system, this probability is neither predictive nor
retrodictive, but a time symmetric inference.) If we discuss so defined
elements of reality, we use this probability in particular in cases
where the intermediate measurement actually is {\it not} performed.
Since in general a measurement significantly changes the whole
set-up\footnote{to clarify this point, another look at the de
Broglie--Bohm model is very instructive: If we describe the experiment
of section \ref{sel} {\it with} intermediate measurement, then, of
course, the ABL formula yields the correct probabilities for the
results, and therefore, the particle is always found in path $d$ (not
in path $c$ !). The reason is that after the measurement the waves of
the quantum system are entangled with those of the intermediate
measurement apparatus, and therefore, the description of section
\ref{bohm} is not valid for this case.}, it is very questionable, whether
such a definition can have any ontological meaning. However, since
Aharonov and Reznik in a recent paper \cite{symm} argue against nonlocality
with the time symmetric formalism, this suggests some idea of reality
behind the mathematical description. This underlying ontology could be
very interesting, because the symmetric formulation includes the
possibility of correlations between different times. Therefore, it
shall be worth checking, if this concept of reality could even contradict
causality.
\\
\section*{Acknowledgements}
\indent

We are grateful to Sheldon Goldstein and Euan Squires for
correspondence and conversation respectively and also to the anonymous
referee for raising the point that we have now addressed in
sec.~\ref{meas}.

B. B. would like to thank the
Cusanuswerk that is supporting him in various ways.
\\[6ex]
{\bf Note added in proof:} Since completing this paper, we have become
aware of some particularly interesting work by T. Mor, L. Goldenberg
and L. Vaidman, who have also considered the double interferometer of
fig.~1 in the context of pre- and postselected elements of reality.\\

\begin{thebibliography}{00}
%
\bibitem{abl}Y. Aharonov, P. G. Bergman, and J. L. Lebowitz, Phys. Rev.
{\bf 134}, B1410 (1964).
\bibitem{gen}Y. Aharonov and L. Vaidman, J. Phys. {\bf A 24}, 2315
(1991).
\bibitem{peres}A. Peres, Phys. Lett. {\bf A 194}, 21 (1994).
\bibitem{dgz}D. D\"urr, S. Goldstein, and N. Zangh\'{\i}, J. Stat. Phys.
{\bf 67}, 843 (1992).
\bibitem{arrow}H. D. Zeh, {\it The Physical Basis of the Direction of
Time} (Springer, Berlin, 1989), p. 97.
\bibitem{belin}F. J. Belinfante, {\it Measurements and Time Reversal in
Objective Quantum Theory} (Pergamon Press, Oxford, 1975).
\bibitem{hape}J. J. Halliwell, J. P\'erez-Mercader, and W. H. Zurek
(eds), {\it Physical Origins of Time Asymmetry} (CUP, Cambridge,
1994).
\bibitem{gri}R. B. Griffiths, J. Stat. Phys. {\bf 36}, 219 (1984).
\bibitem{deco}J. J. Halliwell, {\it A Review of the Decoherent
Histories Approach to Quantum Mechanics}. In D. M. Greenberger and A.
Zeilinger (eds), {\it Fundamental Problems in Quantum Theory} (New
York Academy of Sciences, New York, 1995).
\bibitem{between}Y. Aharonov and L. Vaidman, Phys. Rev. {\bf A 41}, 11
(1990).
\bibitem{peace}A. Shimony, {\it Search for a Worldview Which Can
Accommodate Our Knowledge of Microphysics}. In J. Cushing and E.
McMullin (eds), {\it Philosophical Consequences of Quantum Theory:
Reflections on Bell's Theorem} (University of Notre Dame Press, Notre
Dame, 1989), pp 25 -- 37.
\bibitem{real}M. Redhead, {\it Incompleteness, Nonlocality, and
Realism} (Clarendon, Oxford, 1987), p. 72.
\bibitem{mreal}L. Vaidman, Phys. Rev. Lett. {\bf 70}, 3369 (1993).
\bibitem{neu}J. v. Neumann, {\it Mathematische Grundlagen der
Quantenmechanik} (Springer, Berlin, 1932).
\bibitem{bohm}D. Bohm, Phys. Rev. {\bf 85}, 166, 180 (1952).
\bibitem{tra}C. Dewdney, Phys. Lett. {\bf 109A}, 377 (1985).
\bibitem{empty}L. Hardy, Phys. Lett {\bf A 167}, 11 (1992).
\bibitem{hist}E. J. Squires, in M. Bitbol and O. Darrigol (eds), {\it
E. Schr\"odinger, Philosophy and the Birth of Quantum Mechanics}
(Editions Frontieres, Gif-sur-Yvette, 1993).
\bibitem{symm}Y. Aharonov and B. Reznik, {\it ``On a Time Symmetric
Formulation of Quantum Mechanics''}, TAUP 2200-94, QUANT-PH-9501011.
%
\end{thebibliography}
\end{document}